\documentclass[aps,prl,twocolumn,amsmath,amsfonts,floatfix]{revtex4-1}
\usepackage{bm}
\usepackage{amsmath}
\usepackage{latexsym}
\usepackage{color}
\usepackage{tensor}
\usepackage{braket}
\usepackage{graphicx}
\usepackage{siunitx}
\usepackage{slashed}
\usepackage{dcolumn}   

\newcommand{\nn}{\nonumber}

\begin{document}

\title{Deuteron charge radius from the Lamb-shift measurement in muonic deuterium}

\author{Marcin Kalinowski}
\email[]{mj.kalinowski@student.uw.edu.pl}
\affiliation{Faculty of Physics,
	University of Warsaw,
	Pasteura 5, 02-093 Warsaw, Poland}

\date{\today}

\begin{abstract}
	The deuteron charge radius is calculated from the measurement of the Lamb shift in muonic deuterium, taking into account the electron vacuum polarization correction to the nuclear-structure effects. This correction is unexpectedly large and gives a mean-square charge-radii difference $ r_d^2-r_p^2 = \SI{3.81747\pm 0.00346}{\femto\meter\squared} $, which is now consistent with that obtained from the ordinary H-D isotope shift in the 1$ S $-2$ S $ transition. This suggests that the long-standing discrepancy in the proton charge radius obtained from electronic and muonic systems is due to an underestimated uncertainty in ordinary hydrogen spectroscopy.
\end{abstract}

\maketitle
Atomic measurements are at the frontier of low-energy tests of fundamental interactions, which include the search for electric dipole moment in molecules such as thorium monoxide \cite{acme14}, measurements of parity violation in cesium \cite{wood97}, and the possible dependence of the fundamental constants on time \cite{rosenband08}. So far, none of these methods has indicated any new physics. Recently, an approach based on the comparison of the nuclear charge radius obtained in different ways, such as from muonic and electronic systems, has shown promising results. Due to the very precise theoretical description of the hydrogenic spectra, the charge radius can be extracted from the corresponding spectroscopic experiments. The value of the proton radius $ r_p $ obtained from measurements in muonic hydrogen ($ \mu $H) \cite{pohl2010,antognini_proton_2013}, which is a bound system of the muon and the proton, is in $ 5.6\sigma $ discrepancy with the world-averaged value \cite{mohr_codata_2016} obtained from ordinary hydrogen. Because every relevant contribution in the current theory was taken into account, such disagreement may suggest unknown effects or unknown interactions that could not be explained by a straightforward modification of the standard model \cite{pachucki_review}. This led to extending the study of muonic systems to more complex nuclei, such as muonic deuterium ($ \mu $D) \cite{pohl:16} and helium ($ \mu $He) \cite{pohl_private}. In the case of deuteron, the charge radius $ r_d $ inferred from muonic measurements also deviates by $ 5.6\sigma $ from the CODATA14 world-averaged value \cite{mohr_codata_2016} obtained in ordinary deuterium, and by $ 3.5\sigma $ from the radius extracted in the recent analysis \cite{pohl_deuteron_2017} of spectroscopy measurements in ordinary deuterium only. 

Because the determination of the deuteron charge radius depends on the proton charge radius, the discrepancy in the $ r_p $ affects results for $ r_d $. Therefore, we think that a better way to compare electronic and muonic systems is to combine the results for $ \mu $D and $ \mu $H into a mean-square charge-radii difference $ r_d^2-r_p^2 $ that can be matched against the similar value inferred from very precise measurements \cite{parthey10_1s2s} of the ordinary H-D isotope shift in the 1$ S $-2$ S $ transition. In this approach, the proton contribution cancels out and the difference depends mostly on the deuteron structure radius. According to the latest estimate \cite{hernandez18_deutradpuzz}, the mean-square charge-radii difference $ r_d^2-r_p^2 $ deviates by $ 2\sigma $ between muonic and electronic systems.

Several recent experiments in ordinary hydrogen \cite{beyer17_2s4p,hessels18} favor the smaller proton size and agree with muonic measurements \cite{pohl2010,antognini_proton_2013}, which seems to resolve the discrepancy. In this work we show that, in the case of deuteron, incorporating a missing theoretical contribution resolves the $ 2\sigma $ discrepancy mentioned above.  Natural units ($\hbar = c = \varepsilon_0 = 1$) are used throughout.

Theoretical prediction of the 2$P _{1/2} $-2$ S_{1/2} $ splitting, known as the Lamb shift, in muonic deuterium can be expressed, following Ref.~\cite{krauth:16}, as the sum of the precisely calculated QED contribution \cite{borie12,martynenko11} in the point-nucleus limit, the part proportional to the mean square charge radius $ r_d^2 $ of the deuteron \cite{borie12,karshenboim12,martynenko14} and the nuclear polarizability contribution $ \Delta E_{\rm pol} $ \cite{pachucki11_munucld,pachucki15_munucld,friar13,carlson14,hernandez14}, with the total splitting expressed as
\begin{align}\label{eq:Elsth}
E_{\rm LS} = &\, \SI{228.7766(10)}{\milli\electronvolt}+\Delta E_{\rm pol} \nn\\
&-6.11025(28)r_d^2\, \SI{}{\milli\electronvolt\per\femto\meter\squared},
\end{align}
where $ \Delta E_{\rm pol} $ is the main limiting factor in the precise theoretical description. Nuclear polarizability can be split into two terms
\begin{equation}\label{eq:Epoleq}
\Delta {E}_{\rm pol}^{\rm th} = \delta_{\rm TPE} {E}_{\rm pol}+\delta_{\rm HO} {E}_{\rm pol},
\end{equation}
where $ \delta_{\rm TPE} {E}_{\rm pol} $ contains terms from the two-photon exchange, which are of fifth order in the fine-structure constant $ \alpha $, and additionally the Coulomb distortion correction. According to the latest analysis \cite{hernandez18_deutradpuzz} this part amounts to
\begin{equation}\label{eq:Epolth}
\delta_{\rm TPE} E_{\rm pol} = \SI{1.715(23)}{\milli\electronvolt}
\end{equation}
However, recent calculations \cite{tomalak18_ls} of the nucleon polarizability alter this value. Previously, the authors in Ref.~\cite{hernandez18_deutradpuzz}, following Ref.~\cite{krauth:16}, assumed that single-nucleon interactions amount to $ \SI{0.0471\pm 0.0101}{\milli\electronvolt} $. On the other hand, based on dispersive calculations in Ref.~\cite{tomalak18_ls} we obtained, through the proper scaling,
\begin{equation}\label{eq:E1nucl}
\delta E_{\rm 1 nucl} = \frac{m_r^3(\mu {\rm D})}{m_r^3(\mu {\rm H})}\frac{\SI{0.3066(287)}{\milli\electronvolt}}{8} = \SI{0.0448\pm 0.0042}{\milli\electronvolt},
\end{equation}
which is similar but more than twice as accurate. The reduced muon-nucleus mass is given by \begin{equation}\label{eq:mr}
m_r(\mu{\rm N}) = \frac{m_\mu m_N}{m_\mu+m_N},
\end{equation}
where $ m_\mu $ is the muon mass and $ m_N $ denotes the mass of the appropriate nucleus. Henceforth $ m_r \equiv m_r(\mu{\rm D}) $.

Replacing the single-nucleon interaction contribution to Eq.~\eqref{eq:Epolth} with the result of Eq.~\eqref{eq:E1nucl} gives the new value of the $ \alpha^5 $ two-photon exchange correction,
\begin{equation}\label{eq:Etpe}
\delta_{\rm TPE} E_{\rm pol} = \SI{1.713\pm 0.021}{\milli\electronvolt}.
\end{equation}
All contributions, excluding the Coulomb distortion correction, that are of higher order than $\alpha^5$ constitute $ \delta_{\rm HO} {E}_{\rm pol} $. They were not included in the calculation of $ \Delta E_{\rm pol}^{\rm th} $ in Refs.~\cite{krauth:16,hernandez18_deutradpuzz}, and the only higher-order contribution that has been calculated is the three-photon exchange \cite{pachucki_three-photon-exchange_2018}. Unfortunately, its value is too small to resolve the $ 2\sigma $ discrepancy. 

We report the calculation of the missing contribution, which comes from the unexpectedly large electron vacuum polarization (eVP) correction to the dominant nuclear-structure term.

The leading nuclear polarizability correction is described by the two-photon exchange between the muon and the nucleus. The dominating term comes from the nonrelativistic limit, where, because the distance from the proton to the nuclear center of mass is very small compared to that of the muon, the leading contribution comes from the electric dipole excitations
\begin{align}\label{eq:Epol}
\delta E =& \bigg\langle\psi\,\phi_N\bigg|\vec{R}\cdot\vec{\nabla}\left(\frac{\alpha}{r}\right)\nn\\
&\times\frac{1}{E_N+E_0-H_N-H_0}\vec{R}\cdot\vec{\nabla}\left(\frac{\alpha}{r}\right)\bigg|\psi\,\phi_N\bigg\rangle,
\end{align}
where $ H_0=p^2/(2m_r)-\alpha/r $ is the nonrelativistic Coulomb Hamiltonian for the muon with reduced mass $ m_r $, $ H_N $ is the deuteron Hamiltonian, $ \vec{R} $ is the position of the proton with respect to the nuclear center of mass, $ \psi $ is the muon wave function, and $ \phi_N $ is the nuclear wave function. All values of the fundamental physical constants are from Ref.~\cite{mohr_codata_2016}.

The average nuclear excitation energy $ E $ is much larger than the atomic one, so we perform expansion in the large parameter $ E/(m_r\, \alpha^2)  $ in Eq.~\eqref{eq:Epol}. The leading term is the dipole polarizability
\begin{equation}\label{eq:E0}
\delta E_{\rm pol0} = \frac{4\pi\alpha^2}{3}\psi^2(0)\int_{E_T} d E\, \sqrt{\frac{2m_r}{E}}\,\lvert\braket{\phi_N|\vec{R}|E}\rvert^2.
\end{equation}
It contributes to the Lamb shift by $ \delta E_{\rm pol0}= \SI{1.910}{\milli\electronvolt} $, which is at least an order of magnitude larger than any other nuclear-structure effect (see Table I in Ref.~\cite{pachucki15_munucld}). Therefore, we considered the eVP correction $ \delta_{\rm vp} E_{\rm pol} $ only to this dominating term.

The leading electron vacuum polarization correction $ \delta_{\rm vp}E_{\rm pol} $ to Eq.~\eqref{eq:Epol} is of the order $\alpha^6 $ and has two components, $ \delta_{\rm pot}E_{\rm pol}  $ and $ \delta_{\rm wf}E_{\rm pol} $. The first corresponds to the modification of the photon propagator, which effectively replaces one of the Coulomb potentials $ V=-\alpha/r $ with the term $ \delta V $ from the Uehling potential \cite{uehling35},
\begin{equation}\label{eq:Uehling}
V_{\rm vp} = V + \delta V = - \frac{\alpha}{r}\left(1+\frac{2\alpha}{3\pi}\int_{1}^{\infty}d\xi\,\rho(\xi)\,e^{-2r\,m_e\xi}\right),
\end{equation}
where $ \rho(\xi) $ is a dimensionless function
\begin{equation}\label{eq:rho}
\rho(\xi) = \sqrt{\xi^2-1}\, \frac{2\xi^2+1}{2\xi^4}.
\end{equation}
Neglecting the Coulomb distortion and deuteron quadrupole moment, and approximating $ \psi(r) $ with $\psi(0) $, the leading correction in $ \alpha $ is expressed as
\begin{align}\label{eq:Epot1}
\delta_{\rm pot} E_{\rm pol} =& 2\frac{4m_r\alpha^3}{9\pi}\psi^2(0)\int_{E_T} d E\,\lvert\braket{\phi_N|\vec{R}|E}\rvert^2\int_{1}^{\infty}d\xi\,\rho(\xi) \nn\\
&\times \int \frac{d^3p}{(2\pi)^3}\frac{4\pi}{p^2+4\, m_e^2\, \xi^2}\frac{4\pi}{p^2+2\, m_r\,E},
\end{align}
where $ E $ denotes the nuclear excitation energy and the combinatorial factor 2 at the beginning is due to the modification of one of the two Coulomb potentials. The result of Eq.~\eqref{eq:Epot1} depends on the large parameter $ \sqrt{\frac{E m_r}{2m_e^2}} \sim 20 $. From the first two terms of the expansion, we obtain
\begin{align}\label{eq:Evp1LS}
\delta_{\rm pot} E_{\rm pol} =& \frac{8\alpha^3}{9}\psi^2(0)\int_{E_T} d E\,\lvert\braket{\phi_N|\vec{R}|E}\rvert^2\sqrt{\frac{2 m_r}{E}} \\
&\times\left[\ln\left(\frac{E}{2 m_r}\right)+2\ln\frac{2 m_r}{m_e}-\frac{5}{3}+\frac{3\pi\, m_e}{4\,m_r}\sqrt{\frac{2m_r}{E}}\right]\nn
\end{align}
The numerical value, calculated with the AV18 potential \cite{wiringa:95}, is
\begin{equation}\label{eq:Evp1LSn}
\delta_{\rm pot} E_{\rm pol} = \SI{0.0201}{\milli\electronvolt}.
\end{equation}
The second correction $ \delta_{\rm wf} E_{\rm pol} $ is the result of perturbing the muon wave function $ \psi $ in Eq.~\eqref{eq:Epol} with the potential $ \delta V $ defined in Eq.~\eqref{eq:Uehling},
\begin{equation}\label{eq:vppsi}
\tilde{\psi}(0) = -\int d^3r\, G_{2S}(0,\vec{r})\delta V(r)\psi(r),
\end{equation}
where $ G_{2S}(0,\vec{r}) $ is a special case of the reduced Coulomb Green's function, defined as
\begin{equation}\label{eq:rcgreenf}
G_n(\vec{r}_1,\vec{r}_2) = \Braket{\vec{r}_1|\frac{1}{(H_0-E_n)'}|\vec{r}_2},
\end{equation}
where the prime in the denominator denotes the exclusion of the state $ n $ with the corresponding energy $ E_n $.
The explicit form of formula \eqref{eq:rcgreenf} for the 2$ S $ atomic state was derived in Ref.~\cite{pachucki96_theoryLS},
\begin{align}
G_{2S}(0,\vec{r}) &= \frac{\alpha\, m_r^2}{4\pi} \frac{e^{-x/2}}{4\,x}\big(8+12x-26x^2\nn\\
&+2x^3+8(x-2)x\left(\gamma+\ln x\right)\big),
\end{align}
where $ x=m_r\alpha r $. After integration, Eq.~\eqref{eq:vppsi} gives the value of the perturbed wave function of the 2S state at the origin
\begin{equation}
\tilde{\psi}(0) =0.72615\left(\frac{\alpha}{\pi}\right)\, \psi(0).
\end{equation}
The contribution to the Lamb shift is obtained through the substitution $ \psi^2(0) \rightarrow \psi^*(0)\tilde{\psi}(0) $ in Eq.~\eqref{eq:E0},
\begin{equation}\label{eq:Evp2LS}
\delta_{\rm wf} E_{\rm pol} = 2\, \frac{\tilde{\psi}(0)}{\psi(0)}\,\delta E_{\rm pol0} = \SI{0.0064}{\milli\electronvolt},
\end{equation}
where the factor 2 is from the perturbation of the left and right wave functions.

The total electron vacuum polarization correction to the nuclear structure is the sum of terms in Eqs.~\eqref{eq:Evp1LSn} and \eqref{eq:Evp2LS},
\begin{equation}\label{eq:vptot}
\delta_{\rm vp} E_{\rm pol}  = \SI{0.0265(3)}{\milli\electronvolt},
\end{equation}
where, following Ref.~\cite{pachucki15_munucld}, we assign \SI{1}{\percent} uncertainty.
Together with the inelastic three-photon-exchange correction $ \delta_{\rm 3pe} E_{\rm pol} = \SI{0.00875(92)}{\milli\electronvolt} $, from Ref.~\cite{pachucki_three-photon-exchange_2018}, it gives the higher-order part $ \delta_{\rm HO} E_{\rm pol} = \delta_{\rm vp} E_{\rm pol} + \delta_{\rm 3pe} E_{\rm pol} $ of the nuclear polarizability,
\begin{equation}\label{eq:Eho}
\delta_{\rm HO} E_{\rm pol} = \SI{0.03525(97)}{\milli\electronvolt}.
\end{equation}
The total correction, as expressed in Eq.~\eqref{eq:Epoleq}, with the $ \alpha^5 $ contribution from Eq.~\eqref{eq:Etpe} and the higher-order terms from Eq.~\eqref{eq:Eho}, gives
\begin{equation}\label{eq:Epolthnew}
\Delta E_{\rm pol}^{\rm th} = \SI{1.748\pm 0.021}{\milli\electronvolt},
\end{equation}
where most of the uncertainty comes from an insufficient understanding of electromagnetic interactions of nucleons inside the nucleus.

Measurement in muonic deuterium \cite{pohl:16} gives the experimental value of the Lamb shift,
\begin{equation}\label{eq:ELSexp}
E_{\rm LS}^{\rm expt.} = 202.8785(31)_{\rm stat}(14)_{\rm syst}\, {\rm meV}.
\end{equation}
The mean-square charge radius of deuteron is obtained through Eq.~\eqref{eq:Elsth}, with the updated theoretical polarizability from Eq.~\eqref{eq:Epolthnew},
\begin{subequations}
	\begin{equation}\label{eq:rd2}
	r_d^2 = 4.52453(53)_{\rm prot}(346)_{\rm rest}\, {\rm fm}^2,
	\end{equation}
	where $ (53)_{\rm prot} $ is the uncertainty from the proton polarizability only and $ (346)_{\rm rest} $ is the remainder. The new deuteron charge radius is
	\begin{equation}\label{eq:rdnew}
	r_d = \SI{2.12710\pm 0.00082}{\femto\meter}.
	\end{equation}
\end{subequations}

In order to reliably compare this result with electronic measurements, we use the proton charge radius $ r_p = \SI{0.84087(39)}{\femto\meter}  $ inferred from muonic hydrogen experiments \cite{pohl2010,antognini_proton_2013} to calculate the mean-square charge-radii difference,
\begin{equation}\label{eq:rd2diffmu}
 r_d^2(\mu{\rm D})-r_p^2(\mu{\rm H}) =\SI{3.81747\pm 0.00346}{\femto\meter\squared},
\end{equation}
where the dependence on the proton polarizability cancels out, as does the uncertainty from the proton in Eq.~\eqref{eq:rd2}. This result is consistent with the very precise value obtained from the ordinary H-D isotope shift in the 1$ S $-2$ S $ transition \cite{parthey10_1s2s}, which, modified by the three-photon exchange in Ref.~\cite{pachucki_three-photon-exchange_2018}, is
\begin{equation}\label{eq:drnew}
 r_d^2(e{\rm D})-r_p^2(e{\rm H}) = \SI{3.82070(31)}{\femto\meter\squared}.
\end{equation}

Agreement in the deuteron charge radius suggests that we have sufficient knowledge of the nuclear-structure effects in the Lamb shift to perform calculations of the nuclear polarizability contributions in heavier elements, such as $ ^3 $He, $ ^4$He, $ ^6$Li, and $ ^7$Li. We note, however, that the spin-dependent part of the nuclear polarizability is not well understood, which is reflected in the recently observed $ 5\sigma $ discrepancy between the theoretical prediction \cite{kalinowski18} and the experimental measurement \cite{pohl:16} of the 2S hyperfine splitting in muonic deuterium. In general, to reduce the uncertainty further and increase the accuracy of the test, we should better understand electromagnetic interactions of nucleons inside the nucleus.

We note that the electron vacuum polarization is not the only radiative correction to the nuclear-structure effects. Muonic and nuclear self-energy (SE) corrections are present as well, but we argue that they are significantly smaller than the eVP correction in Eq.~\eqref{eq:vptot}. The nuclear SE is not only small, but also it is partially included in the heuristic proton-neutron potential \cite{wiringa:95} and effective electromagnetic moments of the nucleus. The $ \mu $SE is of the order $ \left(\frac{\alpha}{\pi}\right) $ relative to the leading term in Eq.~\eqref{eq:E0} but does not have the $ m_r/m_e $ enhancement, in contrast to the correction discussed in this work. Moreover, in the point nucleus limit, the value of $ \mu $SE is very small (see Table I in Ref.~\cite{krauth:16}) in comparison with the electron vacuum polarization, which is the leading term in the Lamb shift of muonic systems. It indicates that radiative corrections to the nuclear-structure effects, other than the eVP, can be neglected with the current level of precision.  

In summary, we calculated the electron vacuum polarization correction to the leading nuclear polarizability effect in muonic deuterium, which, combined with other recent results \cite{tomalak18_ls,pachucki_three-photon-exchange_2018,hernandez18_deutradpuzz}, gives a new muonic mean-square charge-radii difference. Its value is in agreement with the very precise result from the ordinary H-D isotope shift in the 1$ S $-2$ S $ transition. This consistency is strong evidence for the correctness of measurements in muonic hydrogen and deuterium. Therefore, it suggests that the current disagreement in the determination of the proton charge radius is caused exclusively by underestimated uncertainty in ordinary hydrogen spectroscopy.

\begin{acknowledgments}
        The author would like to thank Krzysztof Pachucki for useful discussions and Oscar Javier Hernandez and Sonia Bacca for correcting our numerical calculations. This work was supported by the National Science Center (Poland)
        Grant No. 2017/27/B/ST2/02459.
\end{acknowledgments}

%

\end{document}